\documentclass[amsthm]{elsart}
\usepackage{natbib,amsmath,amssymb}
\usepackage[pdfhighlight=/P,colorlinks=true,
linkcolor=blue,citecolor=blue,urlcolor=blue]{hyperref}

\journal{Journal of Symbolic Computation}

\newcommand{\Integer}{\mathbb{Z}}
\newcommand{\Field}{\mathbb{K}}
\newcommand{\coef}{\mathop{\rm coef}\nolimits}
\newcommand{\const}{\mathop{\rm const}}
\newcommand{\lord}{\mathop{\rm\mspace{4mu}\underline{\mspace{-4mu}ord\mspace{-4mu}}\mspace{4mu}}}
\newcommand{\ord}{\mathop{\rm ord}}
\newcommand{\im}{\mathop{\rm Im}}
\newcommand{\Eu}{{\sf E}}
\newcommand{\FF}{{\cal F}}
\newenvironment{code}{\begin{list}{}{\setlength{\leftmargin}{1.5em}
\setlength{\listparindent}{0pt}\setlength{\topsep}{1ex}}\small\item}{\end{list}}

\begin{document}

\begin{frontmatter}
\title{Integrability test for evolutionary lattice equations of higher order}
\thanks{This research was supported by grants RFBR 13-01-00402a and
NSh--3139.2014.2.}
\author{V.E. Adler}
\address{L.D. Landau Institute for Theoretical Physics,
Ak.~Semenov str.~1-A,\\ 142432 Chernogolovka, Russian Federation}
\ead{adler@itp.ac.ru}
\date{25 August 2014}
\begin{abstract}
A generalized summation by parts algorithm is presented for solving of
difference equations of the form $T^m(y)-a[u]y=b[u]$ where $T$ denotes the
shift $u_j\to u_{j+1}$. Solvability of such type of equations with respect to
coefficients of formal symmetry (or formal recursion operator) provides a
convenient integrability test for evolutionary differential-difference
equations $u_{,t}=f(u_{-m},\dots,u_m)$. The algorithm is implemented in {\em
Mathematica}.
\end{abstract}
\end{frontmatter}

\section{Introduction}\label{s:intro}

Existence of an infinite set of higher symmetries is a characteristic property
of integrable equations. For a given evolutionary equation
$\partial_t(u)=f[u]$, it implies solvability of the Lax equation
\begin{equation}\label{i.Gt}
 D_t(G)=[f_*,G]
\end{equation}
where $D_t$ denotes evolutionary derivative corresponding to the equation and
$f_*(v)=df[u+\epsilon v]/d\epsilon|_{\epsilon=0}$ is the linearization
operator. The unknown $G$ (called formal symmetry or formal recursion operator)
is a power series with respect to differentiation $D$ in the continuous case or
to automorphism $T$ in the difference case. Solvability of equation
(\ref{i.Gt}) with respect to the coefficients of $G$ provides a sequence of
necessary integrability conditions which can be applied both for testing of a
given equation and for classification of integrable cases among a whole set of
equations under consideration. Additionally, one can use the conditions which
follow from existence of an infinite set of higher order conservation laws.
This approach allowed to solve a number of classification problems for
integrable partial differential equations of the Korteweg--de Vries and the
nonlinear Schr\"odinger type, see e.g. \citet{Sokolov_Shabat_1984,
Mikhailov_Shabat_Yamilov_1987, Mikhailov_Shabat_Sokolov_1991,
Mikhailov_Shabat_1993, Meshkov_Sokolov_2012} and for differential-difference
equations of the Volterra and the Toda lattice type \citep{Yamilov_1983,
Yamilov_2006, Shabat_Yamilov_1991, Adler_Shabat_Yamilov_2000, Adler_2008}.
Problems of symbolic computation of higher symmetries, conservation laws,
recursion operators and Lax pairs were discussed in many papers, see e.g.
\citet{Goktas_Hereman_1999, Hickman_Hereman_2003, Hereman_SSW_2005,
Sokolov_Wolf_2001, Tsuchida_Wolf_2005}; integrability tests based on these
notions were developed e.g. in \citet{Gerdt_Shvachka_Zharkov_1985, Gerdt_1993,
Hereman_Goktas_Colagrosso_Miller_1998}.

The goal of this article is to describe an algorithm which allows to check the
solvability of equation (\ref{i.Gt}) for a given scalar lattice equation of the
form
\begin{equation}\label{i.ut}
 \partial_t(u_n)=f(u_{n-m},\dots,u_{n+m}),\quad n\in\Integer.
\end{equation}
Recall, that the case $m=1$ (equations of Volterra lattice type) was classified
by \citet{Yamilov_1983}. At $m>1$, only few examples of integrable equations
are known at the moment, the Bogoyavlensky lattices \citep{Bogoyavlensky_1991}
being the most well studied ones.

The operator $f_*$ corresponding to equation (\ref{i.ut}) is of the form
$f_*=\sum f^{(j)}T^j$ where $f^{(j)}=\partial_j(f(u_{-m},\dots,u_m))$,
$\partial_j=\partial/\partial u_j$ and solution of equation (\ref{i.Gt}) is
sought as a power series $G=g_kT^k+g_{k-1}T^{k-1}+\dots$. One can easily see
that equation (\ref{i.Gt}) in each order of $T$ is equivalent to a relation of
the form
\begin{equation}\label{i.gj}
 f^{(m)}T^m(g_j)-g_jT^j(f^{(m)})=b_j,\quad j=k,k-1,\dots
\end{equation}
where $b_j$ is computed explicitly if the coefficients $g_k,\dots,g_{j+1}$ are
already known. Therefore, the integrability test for the lattice equation under
consideration amounts to stepwise checking of whether equation (\ref{i.gj}) is
solvable with respect to $g_j$; if not then it is not integrable, if yes then
one have to compute $g_j$ and to go to the next condition. In practice, such
a test turns out to be very effective, although, formally, checking of infinite
number of conditions is needed in order to prove the integrability.

Although this scheme is rather standard, two technical issues should be
mentioned in the case $m>1$ which were not paid enough attention till now.
Firstly, the form of equations (\ref{i.gj}) depends on the degree $k$ of the
series $G$ which is not known in advance. This question does not stand at all
in the continuous case (for the KdV type equations), because the operation of
root extraction $G^{1/k}$ is defined for generic pseudodifferential operator
$G=g_kD^k+g_{k-1}D^{k-1}+\dots$ and it allows us to reduce the study of formal
symmetries to the case $\deg G=1$. In the difference case (\ref{i.ut}) at
$m=1$, it is also possible to refrain from this question, since, according to
\citet{Levi_Yamilov_1997,Yamilov_2006}, the exhaustive classification here is
based on just few simple conditions which can be easily derived under
nonrestrictive assumptions about the orders of higher symmetries and
conservation laws. In the case $m>1$, this issue was settled in
\citet{Adler_2014} where it was proved that if equation (\ref{i.Gt}) admits a
solution of any degree $k\ne0$ then it admits as well a solution $G$ of degree
$m$, moreover, one can assume without loss of generality that the positive
parts of $G$ and $f_*$ coincide.

Another issue is related with the algorithm of solving of equation (\ref{i.gj})
itself. If $m=1$ then this equation can be brought, by substitution
$g_j=T^{j-1}(f^{(m)})\cdots f^{(m)}y_j$, to the standard form
\[
 (T-1)(y_j)=\tilde b_j.
\]
The problem of inversion of the total difference operator $T-1$ was addressed
by many authors, both in the theory of integrable equations and in the context
of discrete calculus of variations \citep{Kupershmidt_1985,
Hydon_Mansfield_2004, Mansfield_Quispel_2005}, see also \citet{Olver_1993} for
the parallel theory in the continuous case. In particular, it is well known
that $\Field\oplus\im(T-1)=\ker\Eu$ where $\Field$ denotes the field of
constants and $\Eu=\sum T^{-j}\partial_j$ is the difference Euler operator (or
the variational derivative). The preimage of $T-1$ can be computed by use of
the so-called summation by parts algorithm or by use of the discrete homotopy
operator \citep{Hereman_Deconinck_Poole_2006}.

At $m>1$ we arrive at the inversion problem for slightly more general operators
$T^m-a[u]$. Although the setting is quite natural, I was not able to find any
discussion of this problem in literature. The main result of the article is the
description of an algorithm which allows either to solve an equation
\[
 T^m(y)-ay=b
\]
with given functions $a[u],b[u]$ or to prove that solution does not exist.

An approach which makes use of the formal identity
$(T^m-a)^{-1}=T^{-m}(1+aT^{-m}+(aT^{-m})^2+\dots)$ is considered in section
\ref{s:inversion}. This method is quite simple, but, in practice, it is
applicable only if the coefficients $a,b$ are not too complicated.

Section \ref{s:parts} contains a more effective `generalized summation by parts
algorithm' based on simplification of $a,b$ by a sequence of suitable
substitutions. An implementation of this algorithm in the {\em Mathematica}
programming language is presented in appendix \ref{s:mat-parts}.

A variational meaning of operators $T^m-a$ and generalization of the discrete
homotopy operators remain open questions, very interesting from the theoretical
standpoint, however this approach can hardly give an effective computation
scheme for practical applications.

Section \ref{s:sym} contains few basic notions and facts from the symmetry
approach which are necessary to describe the procedure of testing of a given
lattice equation (\ref{i.ut}). Several simple examples are given in Section
\ref{s:ex} accompanied with a sample code in appendix \ref{s:mat-sym}.

\section{Notations}\label{s:defs}

Let $\FF$ be a differential field of functions depending on finite number of
dynamical variables $u_j$, $j\in\Integer$ and let the shift operator $T$ act on
elements of $\FF$ according to the rule
\[
 T^k(f(u_i,\dots,u_j))=f(u_{i+k},\dots,u_{j+k}).
\]
We will assume that the field of constants $\Field$ is equal to $\mathbb{R}$ or
$\mathbb{C}$. The partial derivatives with respect to dynamical variables will
be denoted $\partial_j=\partial/\partial u_j$, $f^{(j)}=\partial_j(f)$. Note
the identity $T^k\partial_j=\partial_{j+k}T^k$.

The orders of a function $f\in\FF$ are defined as follows:
\begin{gather}
\label{ord}
 \lord f=\min\{j:f^{(j)}\ne0\},\quad
 \ord f=\max\{j:f^{(j)}\ne0\},\quad f\ne\const,\\
\label{ordc}
 \lord f=+\infty,\quad \ord f=-\infty,\quad f=\const.
\end{gather}
We will also use the notation
\[
 J(f)=[\lord(f),\ord(f)],\quad J(\const)=\varnothing.
\]

The following operation (extraction) will be used in the summation by parts
algorithm
\begin{equation}\label{Xk}
 g=X_k(f):\quad g^{(k)}=f^{(k)},\quad J(g)=J(f^{(k)}).
\end{equation}
Informally, it can be described as erasing of additive terms in $f$ which do
not depend on the distinguished variable $u_k$. Function $g$ is defined up to
addition of an arbitrary function of variables $u_j$, $j\in
J(f^{(k)})\setminus\{k\}$. In practice, we will define this operation as
follows:
\begin{equation}\label{Xk'}
 X_k(f)=\int f^{(k)}du_k
\end{equation}
where it is assumed that the integrand is cast in a form which does not contain
variables $u_j$ at $j\not\in J(f^{(k)})$ explicitly and the integration
constant does not depend on these variables as well (which is natural).
Alternatively, one can accept the definition
\begin{equation}\label{Xk''}
 X_k(f)=f|_{u_j=c_j,~j\not\in J(f^{(k)})}
\end{equation}
where $c_j$ are any constants such that expression in the right hand side makes
sense (for instance, if $f$ is a polynomial then we can just set to zero all
unnecessary $u_j$).

\section{Solving of equation $T^m(y)-ay=b$}\label{s:Teq}

Let us consider a difference equation of the form
\begin{equation}\label{Teq}
 T^m(y)-ay=b
\end{equation}
where $m>0$, functions $a,b\in\FF$ are given, function $y\in\FF$ is unknown.
Our goal is to obtain an algorithm which allows either to construct the
solution $y$ or to prove that some relation between coefficients $a,b$ does not
hold which is necessary for the existence of $y$.

The uniqueness of solution (if it exists) depends on the form of the
coefficient $a$. If $a\ne T^m(h)/h$ then the solution is unique (indeed, let
$\tilde y$ be another solution, then $a=T^m(y-\tilde y)/(y-\tilde y)$). If
$a=T^m(h)/h$ ($a=1$, in particular) then the solution is defined up to addition
of a term $\const h$, that is $\ker(T^m-a)=\Field h$.

It follows from equations below that if a solution of equation (\ref{Teq})
exists then it is a rational function of coefficients $a,b$ and functions
obtained from $a,b$ by means of the operators $T$, $\partial_j$ and
substitutions $u_j=\const$. In particular, if $a,b$ are rational functions of
the variables $u_j$ then the solution is rational as well. If $a=\const$ and
$b$ is a polynomial then the solution is polynomial as well.

\subsection{Formal inversion of operator $T^m-a$}\label{s:inversion}

1) {\em Case $a=\const\ne0$.} It is easy to see that if also $b=\const$ then
\[
\begin{array}{ll}
 y=c & \text{at}~~ a=1,~ b=0, \\
 \nexists y & \text{at}~~ a=1,~ b\ne0, \\
 y=b/(1-a)  & \text{at}~~ a\ne1
\end{array}
\]
where $c$ is an arbitrary constant.

Let $b\ne\const$, $\lord b=p_1$ and $\ord b=p_2$. If a solution $y$ exists then
$\lord y=p_1$ and $\ord y=p_2-m$ (in particular, this implies $p_2\ge p_1+m$).
Let us consider the equation
\[
 y-a^rT^{-rm}(y)=T^{-m}(b)+\dots+a^{r-1}T^{-rm}(b),\quad r\ge1
\]
which follows from equation (\ref{Teq}). If $r$ is large enough, such that the
inequality $p_1>p_2-m-rm$ holds, then the arguments of two functions in the
left hand side of the equation belong to disjoint sets. This implies the
formula
\begin{equation}\label{y1}
 y=c+\sum_{s=1}^ra^{s-1}T^{-sm}(b)\big|_{u_j=c_j,~j<p_1},\quad
   r=\Bigl\lfloor\frac{p_2-p_1}{m}\Bigr\rfloor
\end{equation}
where $c_j$ are any constants such that the sum in the right hand side is well
defined (for instance, if $b$ is a polynomial then one can just set $c_j=0$)
and $c$ is an undetermined constant (arbitrary if $a=1$). Thus, if a solution
exists then it is of the form (\ref{y1}) and we only have to make a direct
check by substitution into equation (\ref{Teq}). Notice, that if $a\ne1$ then
the problem of choice of constants can be avoided by use of the formula
\[
 y=1/(1-a^r)\sum_{s=1}^ra^{s-1}T^{-sm}(b)\big|_{u_j=u_{j+rm},~j<p_1}
\]
instead of (\ref{y1}).

2) {\em Case $a=\alpha T^m(h)/h$} is brought to the previous one by the change
$y=h\tilde y$. The question, whether the given coefficient $a$ is of such form,
is answered by investigation of auxiliary equation $T^m(z)-z=\log a-\lambda$
with unknown parameter $\lambda$.
\medskip

3) {\em Case $a\ne\alpha T^m(h)/h$}. Let us differentiate equation (\ref{Teq})
with respect to $u_j$:
\[
 T^m(y^{(j-m)})-ay^{(j)}=a^{(j)}y+b^{(j)}.
\]
Elimination of derivatives of $y$ brings to equation
\begin{equation}\label{sumT}
 \sum_sa_sT^{sm}(a^{(j-sm)}y+b^{(j-sm)})=0
\end{equation}
where the sum contains only finite number of nonvanishing terms corresponding
to the values of $s$ from the interval
\[
 \Bigl\lfloor\frac{j-\max(\ord a,\ord b)}{m}\Bigr\rfloor \le s\le
 \Bigl\lfloor\frac{j-\min(\lord a,\lord b)}{m}\Bigr\rfloor.
\]
The coefficients are defined by the recurrent relation $a_{s-1}=a_sT^{sm}(a)$,
$a_0=1$, that is
\[
 a_s=\prod\limits^0_{k=s+1}T^{km}(a),~~s\le0,\qquad
 a_s=1/\prod\limits^s_{k=1}T^{km}(a),~~s\ge0.
\]
The equation corresponding to $j=j+m$ differs from (\ref{sumT}) just by a
factor. Elimination of $T^{sm}(y)$ by use of equation (\ref{Teq}) allows us to
bring (\ref{sumT}) to the form
\[
 A_jy=B_j,\quad j=0,1,\dots,m-1
\]
where coefficient $A_j=\sum_sT^{sm}(\partial_{j-sm}(\log a))$ does not vanish
at least for one value of $j$. So, function $y$ is found explicitly and we only
have to substitute it into (\ref{Teq}) in order to check whether it is a
solution.

\begin{rem}
Equation (\ref{sumT}) makes sense for $a=\const$ as well. In this case it turns
into the set of equations
\begin{equation}\label{Emb}
 \sum_sa^{-s}T^{sm}(b^{(j-sm)})=0,\quad j=0,1,\dots,m-1
\end{equation}
which serve as necessary solvability conditions for equation (\ref{Teq}). In
particular, at $a=1$, $m=1$ this is the usual condition $\Eu(b)=0$. One can
prove that if $a\ne1$ then conditions (\ref{Emb}) are sufficient as well, that
is, this is an exact definition of $\im(T^m-a)$; if $a=1$ then conditions
(\ref{Emb}) characterize the set $\Field\oplus\im(T^m-1)$.
\end{rem}

\subsection{Generalized summation by parts algorithm}\label{s:parts}

The above approach is rather clear and can be easily realized in the computer
algebra systems. Unfortunately, computing of sums in (\ref{y1}) or (\ref{sumT}) is
not too effective in practice. An alternative algorithm makes use of a sequence
of suitable substitutions of the form
\begin{equation}\label{anew}
 y=\tilde y/A,\quad \tilde a=aT^m(A)/A,\quad \tilde b=T^m(A)b
\end{equation}
or
\begin{equation}\label{bnew}
 y=\tilde y-B,\quad \tilde a=a,\quad \tilde b=b+(T^m-a)(B)
\end{equation}
which bring (\ref{Teq}) to equivalent equations of the same type, but with
coefficients which depend on a reduced set of variables. As a result of a
finite number of steps, one can either construct the solution $y$ explicitly or
prove that it does not exist. The substitutions can be applied in different
ways, but the final answer does not depend on this, because of their
reversibility.

The flow of control can be organized by use of inequalities involving the
orders
\[
 \lord a=q_1,\quad \ord a=q_2,\quad \lord b=p_1,\quad \ord b=p_2.
\]
Some conditions imply that substitutions (\ref{anew}), (\ref{bnew}) with
required properties exist, with functions $A,B$ defined from $a,b$ by the
extraction operation (\ref{Xk}). Other conditions mean that equation
(\ref{Teq}) is unsolvable, in such a case the algorithm should return some
nonzero expression which plays the role of an obstacle for existence of a
solution. Analysis of such obstacles is important in a situation when the
equation coefficients contain arbitrary parameters.
\medskip

1) {\em Case $a=\const\ne0$.} If a solution $y$ exists then $J(y)=[p_1,p_2-m]$.
If the inequality
\[
 p_1>p_2-m
\]
holds then solution may be only constant, that is $y=b/(1-a)$ if $a\ne1$ and
$y=c$ if $a=1$, and we only have to check it by inspection. Notice, that the
arbitrary constant $c$ here is the only source of possible nonuniqueness for
the whole algorithm, upon all substitutions below.

Let $p_1\le p_2-m$, then differentiation of equation (\ref{Teq}) with respect
to $u_{p_2}$ yields the relation $b^{(p_2)}=T^m(y^{(p_2-m)})$ which implies the
inequality
\[
 r=\lord b^{(p_2)}\ge p_1+m.
\]
If it fails then the solution does not exist and expression $b^{(r,p_2)}$ is
returned as an obstacle for its existence. If the inequality is fulfilled then
$J(b^{(p_2)})\subseteq[p_1+m,p_2]$ and the change
\[
 B=X_{p_2}(b),\quad y=\tilde y+T^{-m}(B),\quad \tilde b=b-B+aT^{-m}(B)
\]
brings (\ref{Teq}) to an equivalent equation with $J(\tilde
b)\subseteq[p_1,p_2-1]$ (possibly, with $J(\tilde b)=\varnothing$, that is
$\tilde b=\const$).

By repeating this argument while possible (no more than $p_2-p_1+1$ times), we
will either construct the solution as a finite sum $y=T^{-m}(B+\tilde B+\dots)$
or prove that it does not exist.
\smallskip

2) {\em Reduction of the coefficient $a$.} Now let $a\ne\const$. If
\begin{equation}\label{cond1}
 q_1\le q_2-m\qquad \text{and}\qquad \ord\partial_{q_1}(\log a)\le q_2-m
\end{equation}
(possibly $\partial_{q_1}(\log a)=\const$) then $a$ is of the form
\[
 a=A(u_{q_1},\dots,u_{q_2-m})\hat a(u_{q_1+1},\dots,u_{q_2}),\quad
 A=\exp(X_{q_1}(\log a)).
\]
Then the substitution
\[
 y=\tilde y/A,\quad \tilde a=aT^m(A)/A,\quad \tilde b=T^m(A)b
\]
reduces (\ref{Teq}) to an equivalent equation with $J(\tilde
a)\subseteq[q_1+1,q_2]$ (possibly with $\tilde a=\const$). Iteration of this
transformation leads either to the case 1) or to the case when one of
inequalities (\ref{cond1}) fails, that is
\begin{equation}\label{cond2}
 \max(q_1,\ord\partial_{q_1}(\log a))>q_2-m.
\end{equation}
We will assume that this condition is fulfilled from now on. Further
substitutions will not change $a$.
\smallskip

3) {\em Reduction of the coefficient $b$.} First, if $p_2>q_2$ then iteration
of the substitution
\begin{equation}\label{subst}
 B=X_{p_2}(b),\quad y=\tilde y+T^{-m}(B),\quad \tilde b=b-B+aT^{-m}(B)
\end{equation}
brings the problem to the case $p_2\le q_2$. Notice, that instead of this
change we can apply a simpler one
\[
 y=\tilde y+T^{-m}(b),\quad \tilde b=aT^{-m}(b)
\]
with the same effect. However, (\ref{subst}) turns out to be more effective,
because it is desirable to drop the lower order $p_1$ not too much.

Next, if $p_1<q_1$ and a solution $y$ exists then $\ord y\le q_2-m$ and
$-y^{(p_1)}=b^{(p_1)}/a$ from where it follows
\[
 r=\ord(b^{(p_1)}/a)\le q_2-m.
\]
If this inequality fails then the equation does not have a solution and
expression $\partial_r(b^{(p_1)}/a)$ is returned as an obstacle. If the
inequality holds then $J(b^{(p_1)}/a)\subseteq[p_1,q_2-m]$ and therefore the
substitution
\[
 B=X_{p_1}(b/a),\quad y=\tilde y-B,\quad \tilde b=b+(T^m-a)(B)
\]
brings to an equivalent equation with $J(\tilde b)\subseteq[p_1+1,q_2]$.
Iterating of this change brings the problem to the following case 4).
\smallskip

4) {\em Solving of a linear system.} The problem is reduced now to the case
$J(b)\subseteq J(a)=[q_1,q_2]$. If a solution $y$ exists then $J(y)\subseteq
[q_1,q_2-m]$. Therefore, if $q_1>q_2-m$ then equation (\ref{Teq}) may admit
only a constant solution $y=b/(1-a)$ which can be checked by inspection. Let
$q_1\le q_2-m$. Then, according to (\ref{cond2}),
\[
 r=\ord\partial_{q_1}(\log a)>q_2-m\ge\ord y
\]
and differentiation of equation (\ref{Teq}) yields a system of linear equations
with nonzero determinant with respect to $y$, $y^{(q_1)}$:
\[
 \begin{array}{llll}
 a^{(q_1)}y   &+& ay^{(q_1)}       &= -b^{(q_1)},\\
 a^{(q_1,r)}y &+& a^{(r)}y^{(q_1)} &= -b^{(q_1,r)}.
 \end{array}
\]
From here, function $y$ is uniquely determined and, again, we only have to make
a direct check whether it solves equation (\ref{Teq}).

\section{Formal symmetry test}\label{s:sym}

Let us recall some basic notions of the symmetry approach in application to the
scalar evolutionary lattice equations
\[
 \partial_t(u_n)=f(u_{n-m},\dots,u_{n+m}),\quad n\in\Integer
\]
or, in a shorthand notation,
\begin{equation}\label{ut}
 u_{,t}=f(u_{-m},\dots,u_m).
\end{equation}
A detailed exposition can be found in \citet{Mikhailov_Shabat_Yamilov_1987,
Mikhailov_Shabat_Sokolov_1991, Levi_Yamilov_1997, Yamilov_2006}. For any
function $f\in\FF$, the infinite-dimensional vector field
\[
 D_t=\nabla_f=\sum_{j\in\Integer}T^j(f)\partial_j
\]
is called evolutionary derivative and the difference operator
\[
 f_*=\sum_{j\in\Integer}f^{(j)}T^j
\]
is called linearization operator. Differentiation $D_t(g)$ in virtue of
equation (\ref{ut}) is defined, for a function $g\in\FF$, by two equivalent
formulas $D_t(g)=\nabla_f(g)=g_*(f)$.

A lattice equation
\begin{equation}\label{utau}
 u_{,\tau}=g(u_{-l},\dots,u_l)
\end{equation}
is called (generalized) symmetry of equation (\ref{ut}) if differentiations
$D_t,D_\tau$ commute, that is the equality
\begin{equation}\label{fg}
 \nabla_f(g)=\nabla_g(f)
\end{equation}
holds identically with respect to $u_j$. Equation (\ref{ut}) is considered
integrable if it admits symmetries of order $l$ arbitrarily large. Equation
(\ref{fg}) yields, upon the linearization, a more convenient operator equation
\[
 \nabla_f(g_*)=\nabla_g(f_*)+[f_*,g_*].
\]
The degree $m$ of the operator $f_*$ is fixed and this allows us to consider
$g_*$ as an approximate solution of the Lax equation
\begin{equation}\label{Gt}
 D_t(G)=[f_*,G].
\end{equation}
More precisely, it can be proved that existence of a sequence of symmetries of
arbitrarily large orders implies that (\ref{Gt}) admits a solution in the form
of power series
\[
 G=g_kT^k+\dots+g_1T+g_0+g_{-1}T^{-1}+\ldots,\quad g_j\in\FF,\quad k>0
\]
which is called formal symmetry, or formal recursion operator, of lattice
equation (\ref{ut}). Conditions of solvability of equation (\ref{Gt}) with
respect to the coefficients $g_j$ serve therefore as necessary integrability
conditions for equation (\ref{ut}) under consideration. A weak point here is
that the degree $k$ is not known in advance. In the continuous setting, we can
assume that $k=1$ without loss of generality, due to the extraction of root
$G\to G^{1/k}$ which is correctly defined for generic pseudodifferential
operators $G=g_kD^k+\dots+g_1D+g_0+g_{-1}D^{-1}+\ldots$, but in the difference
situation this argument does not work. Nevertheless, it turns out that the
degree $k$ can always be chosen equal to the order $m$ of equation (\ref{ut})
itself (this degree may be not minimal).

\begin{thm}[\citealp{Adler_2014}]\label{th:GG}
If lattice equation (\ref{ut}) admits symmetries (\ref{utau}) of arbitrarily
large order then the Lax equation (\ref{Gt}) admits a solution of the form
\begin{equation}\label{G}
 G=f^{(m)}T^m+\dots+f^{(1)}T+g_0+g_{-1}T^{-1}+\ldots\in\FF((T^{-1})).
\end{equation}
\end{thm}

Now, equation (\ref{Gt}) turns into a convenient and effective test, since the
resulting necessary integrability conditions do not depend on actual orders of
higher symmetries and can be written down intermediately from the right hand
side of equation (\ref{ut}). It is easy to see that collecting of terms with
$T^j$ in (\ref{Gt}) brings to a sequence of recurrent equations of type
(\ref{Teq}) with respect to $g_j$:
\begin{equation}\label{gj}
 T^m(g_j)-a_jg_j=b_j,\quad a_j=\frac{T^j(f^{(m)})}{f^{(m)}},\quad
 j=0,-1,-2,\dots
\end{equation}
where expression
\begin{equation}\label{bj}
 b_j=\frac{1}{f^{(m)}}\Bigl(D_t(g_{j+m})
  -\sum^{m-1}_{s=-m}f^{(s)}T^s(g_{j+m-s})-g_{j+m-s}T^{j+m-s}(f^{(s)})\Bigr)
\end{equation}
involves only coefficients $g_m=f^{(m)},\dots,g_1=f^{(1)}$ which play the role
of initial conditions and coefficients $g_0,\dots,g_{j+1}$ which are already
computed. Thus, the integrability test amounts to step by step checking of
solvability of equations (\ref{gj}).

It can be proved that existence of a symmetry of order $l\ge m+r$ implies that
first $r$ equations (\ref{gj}) can be resolved with respect to
$g_0,\dots,g_{-r+1}$. Symmetries of orders $l\le m$ give no conditions in this
approach, being lost on the background of the trivial symmetry $u_{,\tau}=f$.
Concerning the sufficiency, the fulfilment of first $r$ conditions (\ref{gj})
does not formally guarantee existence of even one generalized symmetry,
however, if $r$ is large enough then it is a very strong evidence of
integrability.

\begin{rem}
In addition to conditions (\ref{gj}), (\ref{bj}) there is a complementary
sequence corresponding to the formal symmetry of the form
\[
 \bar G=f^{(-m)}T^{-m}+\dots+f^{(-1)}T^{-1}+\bar g_0+\bar g_1T+\ldots\in\FF((T)).
\]
Solutions $G,\bar G$ turn out to be equivalent if equation (\ref{ut}) admits a
sequence of conservation laws $D_t(\rho)=(T-1)(\sigma)$ of orders arbitrarily
large. In this case, equation
\begin{equation}\label{R}
 D_t(R)+f_*^\dag R+Rf_*=0
\end{equation}
is solvable and admits a solution of the form
\[
 R=r_lT^l+r_{l-1}T^{l-1}+\ldots~\in\FF((T^{-1})),\quad 0\le l<m,
\]
such that $\bar G^\dag=-RGR^{-1}$ where $\dag$ denotes the conjugation
$(aT^j)^\dag=T^{-j}a$. Formal symmetries $G,\bar G$ can be considered in more
general situation for equations with different negative and positive orders
\[
 u_{,t}=f(u_{-\bar m},\dots,u_m),
\]
however equation (\ref{R}) may admit nonzero solutions only in the symmetric
case $\bar m=m$. It is clear that equations for $\bar g_j$ and $r_j$ are
similar to (\ref{gj}) and can be checked analogously, so we will not discuss
these additional conditions any more.
\end{rem}

It is worth to notice that integrability conditions become especially simple at
$m=1$, that is for the Volterra type lattice equations. It was already
mentioned in Introduction that equations (\ref{gj}) can be brought in this case
to the standard form $(T-1)(y_j)=\tilde b_j$. Moreover, there exists a more
complicated, but still invertible substitution which allows to rewrite these
conditions in the form of conservation laws (possibly, trivial)
\begin{equation}\label{rs}
 D_t(\rho_j)=(T-1)(\sigma_j),\quad j\ge0
\end{equation}
where the so-called canonical densities $\rho_j$, $j>0$ are equivalent to
$j^{-1}\coef_{T^0}G^j$ modulo $\im(T-1)$. Although this form makes no essential
advantage when testing a given equation, it clarifies a general structure of
the integrability conditions.

\begin{prop}
If $m=1$ then solvability of equations (\ref{gj}), (\ref{bj}) is equivalent to
solvability with respect to $\sigma_j\in\FF$ of conservation laws (\ref{rs})
where densities $\rho_j$ are defined by recurrent relations
\begin{gather*}
 \rho_0=\log f^{(1)},\quad \rho_1=f^{(0)}+\sigma_0,\\
 P_{j+1}[-\rho]+f^{(-1)}T^{-1}(f^{(1)}P_{j-1}[\rho])+\sigma_j=0,\quad j>0
\end{gather*}
with polynomials $P_j$ defined by the generating function
\[
 P_0[\rho]+P_1[\rho]\lambda+P_2[\rho]\lambda^2+\ldots
  =\exp(\rho_1\lambda+\rho_2\lambda^2+\rho_3\lambda^3+\ldots).
\]
\end{prop}

The proof can be found in \citet{Adler_2014}. Several first polynomials $P_j$
are
\[
 P_0=1,\quad P_1=\rho_1,\quad P_2=\rho_2+\frac{\rho^2_1}{2},\quad
 P_3=\rho_3+\rho_1\rho_2+\frac{\rho^3_1}{6},~~ \dots
\]
and the corresponding conserved densities are
\begin{align*}
 \rho_2&=f_{-1}T^{-1}(f_1)+\frac{1}{2}\rho_1^2+\sigma_1,\\
 \rho_3&=f_{-1}T^{-1}(f_1\rho_1)+\rho_1\rho_2-\frac{1}{6}\rho_1^3+\sigma_2,\\
 \rho_4&=f_{-1}T^{-1}(f_1(\rho_2+\frac{1}{2}\rho_1^2))
    +\rho_1\rho_3+\frac{1}{2}\rho^2_2-\frac{1}{2}\rho_1^2\rho_2
    +\frac{1}{24}\rho_1^4+\sigma_3.
\end{align*}

In the general case $m>1$, only part of conditions (\ref{gj}) can be rewritten
as conservation laws. For instance, it is easy to prove that if equations
(\ref{gj}) are solvable till $j=-m$ then functions $\sigma,\sigma_1\in\FF$
exist such that
\[
 D_t(\log f^{(m)})=(T^m-1)(\sigma),\quad D_t(f^{(0)}+\sigma)=(T-1)(\sigma_1).
\]

\section{Examples}\label{s:ex}

Here we present several simple examples, in order to clarify various
computational aspects rather than to obtain new results.

\begin{exmp}\label{ex:V}
Solving of equations (\ref{gj}), (\ref{bj}) for the Volterra lattice
\[
 u_{,t}=u(u_1-u_{-1})
\]
and setting all integration constants to zero yields
\[
 g_1=u,\quad g_0=u+u_1,\quad g_{-1}=\frac{uu_1}{u_{-1}},\quad
 g_j=\frac{u(u_1-u_{-1})}{u_j},\quad j<-1.
\]
It is easy to see that the series $G=\sum g_jT^j$ can be rewritten in a closed
form
\[
 G=uT+u+u_1+uT^{-1}+u(u_1-u_{-1})(T-1)^{-1}\frac{1}{u}
\]
which is the well known recursion operator for the Volterra lattice. However,
in most cases expressions for $g_j$ are much more complicated and search of
corresponding recursion operators is a very nontrivial problem. For instance,
in the case of the second order Bogoyavlensky lattice
\[
 u_{,t}=u(u_2+u_1-u_{-1}-u_{-2}),
\]
we get
\begin{gather*}
 g_2=u,\quad g_1=u,\quad g_0=u+u_1+u_2,\quad g_{-1}=0,\\
 g_{-2}=\frac{1}{u_{-2}}(u_{-1}u_1+uu_1+uu_2),\quad
 g_{-3}=-\frac{1}{u_{-3}}(u_{-2}u+u_{-1}u+u_{-1}u_1),\\
 g_{-4}=\frac{1}{u_{-4}u_{-2}}((u_{-3}+u_{-2})u_{-1}u_1+u_{-2}u(u_1+u_2)),
 \quad\dots
\end{gather*}
which gives little hint on the factored form of $G$
\[
 G=u(1+T^{-1}+T^{-2})(T^2u-uT^{-1})(Tu-uT^{-1})^{-1}(Tu-uT^{-2})(u-uT^{-2})^{-1}.
\]
This is a particular example of recursion operators found by \citet{Wang_2012}
for the Bogoyavlensky lattices of any order $m$. Notice, that in these
operators all inverse factors are binomial and therefore computation of $G(f)$
for a given function $f$ amounts to solving of a sequence of equations of the
type (\ref{Teq}).
\end{exmp}

\begin{exmp}\label{ex:class}
As a sample classification problem, consider a Bogoyavlensky type equation
\[
 u_{,t}=u(u_2+k_1u_1+k_2u+k_3u_{-1}+k_4u_{-2})
\]
with undetermined coefficients. Application of the formal symmetry test yields
on the first step the obstacle
\[
 (1+k_2+k_4)u+(k_1+k_3)u_1=0\quad\Rightarrow\quad k_2=-1-k_4,\quad k_3=-k_1.
\]
After the substitutions, $g_0$ is successfully found, but computing of $g_{-1}$
encounters the next obstacle
\[
 k_1(k_1+k_4)(u-u_2)+k_1(k_1-1)(u_{-1}-u_1)=0.
\]
If $k_1\ne0$ then $k_1=1$, $k_4=-1$ and we arrive to the Bogoyavlensky lattice.
If $k_1=0$ then computation of $g_{-2}$ brings to the obstacle
\[
 k_4(1+k_4)/u_{-2}=0
\]
and we get two more integrable (albeit disappointing) cases: a linearizable
equation $u_{,t}=u(u_2-u)$ and the stretched Volterra lattice
$u_{,t}=u(u_2-u_{-2})$.
\end{exmp}

\begin{exmp}\label{ex:V5}
According to \citet{Yamilov_2006}, the lattice equation
\[
 u_{,t}=h(u_1-u)+h(u-u_{-1})
\]
is integrable if $h$ satisfies equation
\begin{equation}\label{h'}
 h'=\alpha h^2+\beta h+\gamma
\end{equation}
with arbitrary constant coefficients. Equation (\ref{h'}) can be solved in
elementary functions, but this leads to consideration of several cases
corresponding to different parameter sets and special solutions. In order to
handle the whole family in a uniform manner we only have to compute the
coefficients $b_j$ (\ref{bj}) modulo a rule which replaces first and second
derivatives of $h$ in virtue of (\ref{h'}). After this, equations (\ref{gj})
are solved as usual by the summation by parts algorithm.
\end{exmp}

\begin{exmp}\label{ex:mB}
In the above examples, equations pass the test for any choice of integration
constants, but this is not always the case. Consider the modified Bogoyavlensky
lattice
\begin{equation}\label{mB}
 u_{,t}=u(u_2u_1-u_{-1}u_{-2})
\end{equation}
with $f^{(2)}=uu_1$. It is easy to see that operator $T^2-T^j(f^{(2)})/f^{(2)}$
possesses nontrivial kernel for any $j$, so that the general solution of
equation (\ref{gj}) contains an arbitrary constant $c_j$ on each step. However,
it turns out that $c_{-2k+1}$ becomes an obstacle when we proceed to computing
of $g_{-2k}$ and, as a result, the test passes only if we set to zero every
second integration constant. This indicates that the minimal degree of the
formal symmetry $G$ is equal to 2, so that (\ref{mB}) cannot be a symmetry of
an equation of order 1.

The same is true for equation
\begin{equation}\label{mV-}
 u_{,t}=(u^2+1)((u_1^2+1)(u_2-u)+(u_{-1}^2+1)(u-u_{-2}))
\end{equation}
which is related by the non-autonomous change $u_n=(-1)^nu_n$ to the second
order symmetry of the modified Volterra lattice
\[
 u_{,\tau}=(u^2+1)(u_1-u_{-1}).
\]
\end{exmp}

\section{Conclusion}\label{s:conc}

The presented algorithm is designed for straightforward computation of the
formal symmetry for a scalar evolutionary lattice equation of any order. It is
suitable mainly for testing integrability of a single equation or a family
depending on several parameters.

Further generalizations may include equations with two or more components such
as the Toda or the Ablowitz--Ladik type lattices and their `hungry' analogs. In
this case the formal symmetry coefficients are matrices and the inversion of
difference operators becomes a more difficult problem. The vectorial case
\citep[see e.g.][]{Adler_2008} can also be handled by enlarging the set of
dynamical variables.

Concerning the classification problem in general, a lot of results were
obtained in the continuous case for scalar evolutionary equations of orders
3,5,7, see \citet{Meshkov_Sokolov_2012} and references therein. Moreover, there
is a conjecture that all (or at least all polynomial) integrable equations of
higher orders are symmetries of equations of orders 3 and 5, so that there is
just a finite set of integrable hierarchies. In the difference case the
classification is much more difficult and the \citet{Yamilov_1983} list of the
first order lattices remains the only rigorous result obtained so far. The
known examples show that there are primitive integrable lattice equations of
any orders, and description of the set of integrable hierarchies is a
challenging problem, even in the polynomial case.

\appendix
\section{{\em Mathematica} implementation of algorithms}\label{s:mat}

\subsection{Generalized summation by parts}\label{s:mat-parts}

Let $f$ be an expression depending on the variables $u_j$. The following lines
define the shift $T^k(f)$, a list of variables involved in $f$, and orders of
(unsimplified form of) $f$:
\begin{code}
\begin{verbatim}
T[f_, k_] := f /. u[j_] :> u[j + k]
vars[f_] := Union[Cases[f, _u, {0, Infinity}]]
ords[f_] := If[# == {}, {Infinity, -Infinity},
        {#[[1, 1]], #[[-1, 1]]}] &[vars[f]]
\end{verbatim}
\end{code}
For the sake of simplicity, we will assume that all expressions under
consideration are rational, then the command \verb|ords[Together[f]]| returns
correct orders (\ref{ord}), (\ref{ordc}) of $f$.

Function \verb|psum[m,a,b]| defined below solves the equation $T^m(y)-ay=b$. It
returns a pair of expressions $(y,z)$ where $z$ (obstacle) vanishes if and only
if the equation is solvable. If this is the case then $y$ is the general
solution of equation, with possible integration constant denoted by the symbol
\verb|const|. The computation is performed according to the algorithm described
in section \ref{s:parts}, with substitutions (\ref{anew}), (\ref{bnew})
realized as recursive calls (function \verb|psum| just blocks the default
limitation on the recursion depth and calls another function which makes all
job). The computation stops either when some necessary condition for existence
of solution fails or when the solution can be found intermediately.
\begin{code}
\begin{verbatim}
psum[m_, a_, b_] /; m > 0 :=
  Block[{$RecursionLimit = Infinity}, psu[m, a, b]]

psu[m_, aa_, bb_] := Module[
    {a = Together[aa], b = Together[bb],
     A, B, p1, p2, q1, q2, r, y},
    q2 = ords[a]; q1 = q2[[1]]; q2 = q2[[2]];
    p2 = ords[b]; p1 = p2[[1]]; p2 = p2[[2]];

    Catch[
      If[a === 0, Throw[{T[b, -m], 0}]];

      (* Case a = const *)
      If[q1 == Infinity,
        If[p2 < p1 + m,
          y = If[a === 1, const, b/(1 - a)];
          Throw[Together[{y, T[y, m] - a y - b}]]
          ];
        B = Together[D[b, u[p2]]];
        r = ords[B][[1]];
        If[r < p1 + m, Throw[{0, D[B, u[r]]}]];
        B = Integrate[B, u[p2]];
        Throw[psu[m, a, b - B + a T[B, -m]] + {T[B, -m], 0}]
        ];

      (* Reduction of a *)
      A = Together[D[a, u[q1]]/a];
      r = ords[A][[2]];
      If[And[q1 <= q2 - m, r <= q2 - m],
        A = Exp[Together[Integrate[A, u[q1]]]];
        Throw[Together[psu[m, a T[A, m]/A, b T[A, m]]/A]]
        ];

      (* Reduction of b *)
      If[p2 > q2,
        B = Integrate[D[b, u[p2]], u[p2]];
        Throw[psu[m, a, b - B + a T[B, -m]] + {T[B, -m], 0}]
        ];

      If[p1 < q1,
        B = Together[D[b, u[p1]]/a];
        r = ords[B][[2]];
        If[r > q2 - m, Throw[{0, D[B, u[r]]}]];
        B = Integrate[B, u[p1]];
        Throw[psu[m, a, b + T[B, m] - a B] - {B, 0}]
        ];

      (* Solving of a linear system *)
      y = If[q1 <= q2 - m,
          -(D[b, u[q1]]D[a, u[r2]] - D[b, u[q1], u[r2]]a)/
            (D[a, u[q1]]D[a, u[r2]] - D[a, u[q1], u[r2]]a),
          b/(1 - a)];
      Throw[Together[{y, T[y, m] - a y - b}]]
      ]
    ]
\end{verbatim}
\end{code}

\subsection{Computation of formal symmetry}\label{s:mat-sym}

The following cell defines the differential $\sum_jf^{(j)}du_j$ and the
evolutionary derivative $\nabla_g(f)=f_*(g)$:
\begin{code}
\begin{verbatim}
df[f_] := Plus @@ (D[f, #]dif[#] & /@ vars[f])
dt[f_, g_] := df[f] /. dif[u[j_]] :> T[g, j]
dt[f_] := dt[f, F]
\end{verbatim}
\end{code}
The global variables \verb|m,F| will be used to denote the order and the right
hand side of the lattice equation $u_{,t}=f[u]$ under consideration. Commands
in the next cell define partial derivatives $f^{(j)}$, the positive part of
formal symmetry $G_{>0}=(f_*)_{>0}$ as initial conditions for further
computation and coefficients of equation (\ref{gj}). Procedure
\verb|mytest[k]| computes coefficients $g_0,\dots,g_{-k}$ while it is
possible and stops if an obstacle occurs.
\begin{code}
\begin{verbatim}
Clear[m, a, b, c, f, g]
f[j_] := D[F, u[j]]
g[j_] /; j > 0 := f[j]
a[j_] := T[f[m], j]/f[m]
b[j_] := 1/f[m](dt[g[j + m]] - Sum[f[s]T[g[j + m - s], s] -
            g[j + m - s]T[f[s], j + m - s], {s, -m, m - 1}])
mytest[k_] := Do[
    ps = Factor[psum[m, a[j], b[j]]];
    obst = ps[[2]];
    Print[obst];
    If[Not[obst === 0], Break[]];
    g[j] = ps[[1]] /. const -> c[j],
    {j, 0, -k, -1}]
\end{verbatim}
\end{code}

{\em Example \ref{ex:V}.} Next cell demonstrates the basic usage of the above
commands by the examples of the Volterra lattice, its second order symmetry and
the Bogoyavlensky lattice.
\begin{code}
\begin{verbatim}
F = u[0](u[1] - u[-1]);
F2 = u[0](u[1](u[2] + u[1] + u[0]) - u[-1](u[0] + u[-1] + u[-2]));
Expand[dt[F, F2] - dt[F2, F]]

m = 1;
mytest[4]
Table[Factor[g[j] /. c[j_] :> 0], {j, m, -4, -1}]

m = 2;
F = F2;
mytest[4]
Table[Factor[g[j] /. c[j_] :> 0], {j, m, -4, -1}]

F = u[0](u[2] + u[1] - u[-1] - u[-2]);
mytest[6]
Table[Factor[g[j] /. c[j_] :> 0], {j, m, -6, -1}]
\end{verbatim}
\end{code}
The output of \verb|mytest| consists here from a sequence of zeroes which means
that the computation encounters no obstacles. The actual coefficients of the
formal symmetry are stored as the variables \verb|g[j]|.
\smallskip

{\em Example \ref{ex:class}.} A sample classification problem solved by
analyzing of the obstacles to the test.
\begin{code}
\begin{verbatim}
m = 2;
F = u[0](u[2] + k1 u[1] + k2 u[0] + k3 u[-1] + k4 u[-2]);
mytest[6]
Collect[-obst, _u]

Out: (1 + k2 + k4) u[0] + (k1 + k3) u[1]

F = u[0](u[2] + k1 u[1] - (1 + k4) u[0] - k1 u[-1] + k4 u[-2]);
mytest[6]
Collect[-obst, _u]

Out: -(-1 + k1) k1 u[-1] - k1 (k1 + k4) u[0] -
     (1 - k1) k1 u[1] - k1 (-k1 - k4) u[2]

F = u[0](u[2] - (1 + k4) u[0] + k4 u[-2]);
mytest[6]
obst

Out: -2 k4 (1 + k4) / u[-2]
\end{verbatim}
\end{code}

{\em Example \ref{ex:V5}.} The following modification of the test includes an
additional transformation rule. It is applied to the lattice equation with
function $h$ in the r.h.s. which is defined as a solution of an ODE.
\begin{code}
\begin{verbatim}
mytest1[k_, ru_] := Do[
    ps = Factor[psum[m, a[j] /. ru, b[j] /. ru]];
    obst = ps[[2]];
    Print[obst];
    If[Not[obst === 0], Break[]];
    g[j] = ps[[1]] /. const -> c[j],
    {j, 0, -k, -1}]
m = 1;
F = h[u[1] - u[0]] + h[u[0] - u[-1]];
P[x_] := \[Alpha] x^2 + \[Beta] x + \[Gamma]
mytest1[4, {h'[x_] :> P[h[x]], h''[x_] :> P'[h[x]]P[h[x]]}]
\end{verbatim}
\end{code}

{\em Example \ref{ex:mB}.} First two calls of the test for the modified
Bogoyavlensky lattice show that integration constants $c_{-1},c_{-3}$ are
obstacles; test passes after setting to zero all constants with odd numbers.
The same is true for equation (\ref{mV-}). In contrast, second order symmetry
of the modified Volterra lattice passes the test for arbitrary integration
constants.
\begin{code}
\begin{verbatim}
m = 2;
F = u[0](u[2]u[1] - u[-1]u[-2]);
mytest[6]
c[-1] := 0
mytest[6]
c[j_] /; OddQ[j] := 0
mytest[6]
Clear[c]

F = (u[0]^2 + 1)((u[1]^2 + 1)(u[2] - u[0]) +
          (u[-1]^2 + 1)(u[0] - u[-2]));
mytest[4]
c[j_] /; OddQ[j] := 0
mytest[4]
Clear[c]

F = (u[0]^2 + 1)((u[1]^2 + 1)(u[2] + u[0]) -
          (u[-1]^2 + 1)(u[0] + u[-2]));
mytest[4]
\end{verbatim}
\end{code}


\end{document}